\documentclass[a4paper,11pt]{article}
\usepackage{pos}
\usepackage{braket}
\usepackage{booktabs}
\usepackage{makecell}
\usepackage{subcaption}
\usepackage{hyperref}

\title{Quantum algorithms for the simulation of perturbative QCD processes}

\author*[a]{Herschel A. Chawdhry}
\author[b]{Mathieu Pellen}

\affiliation[a]{Department of Physics, University of Oxford, Oxford, England}

\affiliation[b]{Albert-Ludwigs-Universit\"at Freiburg, Physikalisches Institut, Freiburg, Germany}

\emailAdd{herschel.chawdhry@physics.ox.ac.uk}
\emailAdd{mathieu.pellen@physik.uni-freiburg.de}

\abstract{Quantum computers are expected to give major speed-ups for the simulation of quantum systems. In these conference proceedings, we discuss quantum algorithms for the simulation of perturbative Quantum Chromodynamics (QCD) processes. In particular, we describe quantum circuits for simulating the colour part of the interactions of quarks and gluons. We implement our circuits on a simulated noiseless quantum computer and validate them by calculating colour factors for various examples of Feynman diagrams.}

\renewcommand{\hookAfterAbstract}{%
\par\bigskip
\textsc{Pre-print numbers}: FR-PHENO-2023-10, OUTP-23-09P
}

\FullConference{16th International Symposium on Radiative Corrections: Applications of Quantum Field Theory to Phenomenology (
RADCOR2023)\\
28th May - 2nd June, 2023\\
Crieff, Scotland, UK\\}

\begin{document}
\maketitle

\section{Introduction}
Perturbative Quantum Chromodynamics (QCD) calculations provide high-precision predictions of the scattering of fundamental particles, especially at hadron colliders, and are therefore a vital part of the Large Hadron Collider (LHC) physics program.
Calculational complexity presents a key limiting factor in producing these predictions and so the development of new computational techniques is central to advancing the state of the art.
In this conference proceedings paper, based on our article~\cite{Chawdhry:2023jks}, we explore whether future quantum computers could help perform perturbative QCD calculations.
In particular, as a first step towards this goal, we focus here on the simulation of colour in perturbative QCD using a quantum computer.

Quantum computing was first proposed 4 decades ago~\cite{Benioff:1979ce,Feynman:1981tf} and has been of great interest over the years because for certain problems it promises large speed-ups.
In particular, it promises exponential speed-ups for prime factorisation~\cite{Shor.365700} and quadratic speed-ups for generic unstructured search problems~\cite{10.1145/237814.237866} (of which Monte Carlo integration is an example).
A further application is the simulation of quantum systems: since quantum computers perform calculations by manipulating the quantum states of a system, it is natural to use a quantum computer to simulate other quantum systems.
In particular, active fields of research exist studying methods to use quantum computers to perform simulations of quantum chemistry~\cite{ChemistryRev,RevModPhys92015003}, condensed matter systems~\cite{RevModPhys.86.153,qute.201900052}, and lattice QCD~\cite{Klco:2021lap,Bauer:2022hpo}.

In contrast to the many proposals in recent years for the quantum simulation of lattice QCD, the quantum simulation of perturbative QCD has largely remained unexplored, with the exception of some work on parton showers~\cite{Bepari:2020xqi,Bauer:2019qxa,Bepari:2021kwv,Gustafson:2022dsq}.
In this work we take the first steps towards the simulation of generic perturbative QCD processes by presenting algorithms for the quantum simulation of colour.
Compared to the kinematic components of QCD calculations, colour is relatively simple but it is still a good starting point since it presents some of the general challenges of using a quantum computer to simulate perturbative QCD.
The colour parts of calculations therefore provide a useful simplified setup in which to develop general techniques, while allowing the results to be verified against analytic expectations.
One should note, however, that for a sufficiently complicated QCD process, even the colour part would become non-trivial to calculate analytically, and in those cases a quantum simulation of colour could be a valuable standalone result.

Research on this topic is timely. Although the idea of quantum computer has been around for 30-40 years with steady incremental progress on the hardware and software sides, recent years have seen notable commercial interest and increased prospects for the emergence of practical machines in the coming years.
In particular, IBM has since 2019 produced a series of quantum computers with several hundred qubits, albeit subject to hardware noise and without full connectivity, and over the next few years the company aims to increase this to several thousand qubits and implement error correction.
Other companies such as Google and Microsoft have also invested in this area and are aiming to produce an error-corrected general-purpose quantum computer within a decade.
In light of this, there have been various applications proposed in the experimental and theoretical branches of high-energy physics~\cite{Agliardi:2022ghn, Bauer:2019qxa, Bauer:2021gup, Bepari:2020xqi, Bepari:2021kwv, Bravo-Prieto:2021ehz, Cervera-Lierta:2017tdt, Chawdhry:2023jks, Clemente:2022nll, Cruz-Martinez:2023vgs, Fedida:2022izl, Gustafson:2022dsq, Kiss:2022pjw, Li:2021kcs, Perez-Salinas:2020nem, Ramirez-Uribe:2021ubp, Rigobello:2023ype, Williams:2023muq,Nicotra:2023rmn,Nagano:2023kge,Turco:2023rmx,Bass:2023hoi,DiMeglio:2023nsa,Bermot:2023kvh,Sborlini:2023uyq,Humble:2022klb,Hayata:2023bgh}.

There are several specific motivations for applying quantum computing to simulate perturbative QCD.
One reason is that perturbative QCD requires the quantum-coherent summation of many contributions (e.g. from many Feynman diagrams), and this is something that quantum computers are naturally suited to do because they are inherently designed to manipulate quantum superpositions of states.
It also follows that QCD processes with high-multiplicity final states, which are currently described by parton showers, could be well-suited to these quantum computing techniques.
More generally, if perturbative QCD processes can be simulated on a quantum computer, then one could subsequently use existing quantum algorithms that are known to give quantum speed-ups, for example quadratic speed-ups when using quantum computers for Monte-Carlo integration.

\section{Quantum circuits for colour}
In this section we will describe quantum circuits for simulating colour in perturbative QCD.
We will work in the quantum circuit model of quantum computing, which is one of the most widely used models.
It is based on the concept of qubits, i.e. two-state quantum systems like spin-half particles, which are represented on a quantum circuit diagram as horizontal lines.
The operations performed on them are called gates, analogously to the \textsc{and} and \textsc{or} gates used in classical computing.
Since quantum mechanical operations are linear, they are represented as matrices.
A single-qubit operation is represented by a 2-by-2 matrix, and in general an operation acting on $n$ qubits at the same time is represented by a $2^n$-by-$2^n$ matrix acting on the $2^n$ basis states of those $n$ qubits.
The matrices must be unitary, since quantum mechanical operators are always unitary.

Let us start by briefly recalling how colour is calculated in QCD.
Given a Feynman diagram, the corresponding term in the amplitude contains a factor $T^a_{ij}$ for each quark-gluon vertex, and a factor $f^{abc}$ for each triple-gluon vertex, where $T^a_{ij}$ are the generators of $\mathfrak{su}(3)$ in the defining representation and $f^{abc}$ are the structure constants of $\mathfrak{su}(3)$.
For example, the quark self-energy diagram shown on the left of Fig.~\ref{fig:quarkselfenergy} has colour factor
\begin{equation}\label{eq:self_energy_colour_factor}
\mathcal{C} = \sum_{
\substack{
a \in \{1, ..., 8\}\\
i,j,k \in \{1,2,3\}
}
} T^a_{ij} T^a_{jk} \delta_{ik},
\end{equation}
where the Feynman rules require us to sum over intermediate states $j \in \{1,2,3\}$ and $a \in \{1, \ldots, 8\}$, and in this case we have further opted to trace over the initial colour $i$ and final colour $k$ of the quark line.

Noting that the generators $T^a_{ij}$ are linear operators (and are by convention written in terms of the Gell-Mann matrices $\lambda^a$ by defining $T^a = \frac{1}{2} \lambda^a$) and that quantum gates are linear operators, it is natural to ask whether the $T^a_{ij}$ can be implemented as quantum gates and hence be used to simulate the colour part quark-gluon interactions.
We will find that the short answer is yes, but there are complications.
One relatively minor complication is that the matrices are not of the form $2^n$-by-$2^n$, required for the reasons stated above.
A second, more important complication is that the Gell-Mann matrices are not unitary (but are instead Hermitian), as can be immediately seen by observing that most of them contain a row that is entirely zero.
Details on the resolution of these issues can be found in our article~\cite{Chawdhry:2023jks}.

The key results of this work are two quantum gates, $Q$ and $G$, which simulate the colour parts of the quark-gluon and triple-gluon interactions respectively.
Our intention is that these gates can then be composed together, matching the factors appearing in a Feynman diagram calculation, and hence simulate the colour part of the perturbative calculation of a scattering process.
In this conference proceeding we will only give a high-level overview of how these gates are used, and refer the interested reader to our article~\cite{Chawdhry:2023jks} for the detailed designs of these gates.

Since each gluon has 8 basis colour states, it can be represented by the $2^3 = 8$ basis states of 3 qubits.
The 3 basis colour states of a quark are represented by 3 of the $2^2 = 4$ basis states of 2 qubits, the 4\textsuperscript{th} state remaining unused.
The $Q$ gate acts on 3 qubits representing a gluon, 2 qubits representing a quark line, and some extra qubits $\mathcal{U}$ (whose purpose will be described later).
If the gluon qubits are in a colour basis state $\ket{a}_g$, where $a \in \{1, \ldots, 8\}$, and the quark qubits are in a colour basis state $\ket{k}_q$, where $k \in \{1,2,3\}$, and if the qubits $\mathcal{U}$ are in a special reference state $\ket{\Omega}_\mathcal{U} \equiv \ket{0\ldots 0}_\mathcal{U}$, then $Q$ acts in the following way:
\begin{equation}\label{eq:Q_gate_behaviour}
Q\ket{a}_g\ket{k}_q\ket{\Omega}_\mathcal{U} = \sum_{j=1}^3 T^a_{jk} \ket{a}_g\ket{j}_q\ket{\Omega}_\mathcal{U} + \left(\textrm{terms orthogonal to } \ket{\Omega}_\mathcal{U} \right) .
\end{equation}
Since quantum gates are linear operators, if the quark qubits (or gluon qubits, or both) are in superpositions of colour basis states, possibly entangled with other qubits in the circuit, then $Q$ acts linearly on each basis component in the superposition.

For the triple-gluon interaction, we have designed a gate $G$ acting on 3 registers, each of which comprises 3 qubits to represent the colour of a gluon as before.
Given 3 gluon registers $g_1$, $g_2$, $g_3$ in colour basis states $\ket{a}_{g_1}$, $\ket{b}_{g_2}$, $\ket{c}_{g_3}$, the $G$ gate acts in the following way:
\begin{equation}\label{eq:G_gate_behaviour}
G \ket{a}_{g_1}\ket{b}_{g_2}\ket{c}_{g_3}\ket{\Omega}_{\mathcal{U}} = f^{abc} \ket{a}_{g_1}\ket{b}_{g_2}\ket{c}_{g_3}\ket{\Omega}_{\mathcal{U}} + \left(\textrm{terms orthogonal to } \ket{\Omega}_\mathcal{U} \right) ,
\end{equation}
where we include the same extra qubits $\mathcal{U}$ as above.
The reason for including $\mathcal{U}$ can now be seen: while multiplying by $f^{abc}$ is not a unitary operation, the inclusion of extra qubits $\mathcal{U}$ allows a unitary operation~\eqref{eq:G_gate_behaviour} to be defined.
We call these extra qubits the \emph{unitarisation register} and in our article~\cite{Chawdhry:2023jks} we give a detailed description of its usage.
For now we just mention that the number of extra qubits is very small (logarithmic in the number of vertices in the Feynman diagram).
We can then interpret eq.~\eqref{eq:G_gate_behaviour} to mean that when projected onto the special reference state $\ket{\Omega}_\mathcal{U}$ of the unitarisation register, the equation simulates the colour interaction of 3 gluons.

\section{Illustrative example}\label{sec:illustrative_example}
\begin{figure}[]
\center
        \begin{subfigure}{0.35\textwidth}
                 \includegraphics[width=\textwidth]{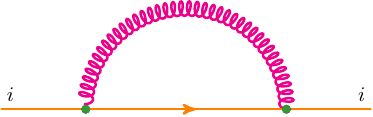}
        \end{subfigure}
\hfill
        \begin{subfigure}{0.60\textwidth}
                 \includegraphics[width=\textwidth]{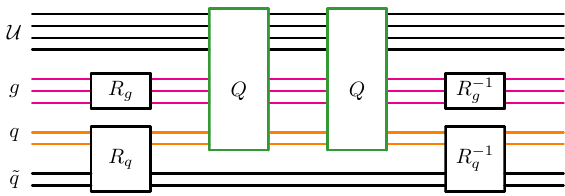}
        \end{subfigure}
        \caption{\label{fig:quarkselfenergy}%
                Example Feynman diagram (left) and a graphical representation of its corresponding circuit (right).}
\end{figure}
We will now work through a simple example in order to illustrate how the $Q$ and $G$ gates can be used.
A generalisation to arbitrarily complicated cases will be given in sec.~\ref{sec:generalisation}.
Consider the Feynman diagram shown in Fig.~\ref{fig:quarkselfenergy}.
It has one quark and one gluon.
As mentioned above, we use 2 qubits to represent the colour of the quark and 3 qubits to represent the colour of the gluon.
There is a complication: in order to be able to compute the trace, we introduce for each quark line a pair of 2-qubit registers $q$ and $\tilde{q}$, rather than just a single 2-qubit register.
The extra register $\tilde{q}$ is not affected by the simulation gates $Q$ or $G$, but instead exists solely to allow the compution of the trace, as will be seen shortly.

We start the circuit in a reference state $\ket{\Omega}_g\ket{\Omega}_q\ket{\Omega}_{\tilde{q}}\ket{\Omega}_\mathcal{U}$, where $\ket{\Omega}_r$ indicates that each qubit of a register $r$ is in the state $\ket{0}$.
We then apply a gate $R_g$ to the gluon register to rotate it into an equal superposition of all 8 basis colour states:
\begin{equation}
\label{eq:Rg_effect}
R_g\ket{\Omega}_g = \sum_{a=1}^8 \frac{1}{\sqrt{8}} \ket{a}_g.
\end{equation}
The explicit form of $R_g$ can be found in the Appendix of ref.~\cite{Chawdhry:2023jks}.
The gate $R_q$ (also defined in the Appendix of ref.~\cite{Chawdhry:2023jks}) is now applied to the quark registers to place them into the following equal superposition of states:
\begin{equation}\label{eq:Rq_effect}
R_q \ket{\Omega}_q \ket{\Omega}_{\tilde{q}} = \sum_{k=1}^3 \frac{1}{\sqrt{3}} \ket{k}_q\ket{k}_{\tilde{q}},
\end{equation}
where it should be observed that the $q$ and $\tilde{q}$ registers are entangled.
Thus, after applying the $R_g$ and $R_q$ gates, the quantum computer is in the state
\begin{equation}
\frac{1}{\sqrt{24}} \sum_{a=1}^8 \sum_{k=1}^3 \ket{a}_g\ket{k}_q\ket{k}_{\tilde{q}}\ket{\Omega}_\mathcal{U}.
\end{equation}

We now perform the key simulation steps, where we apply two $Q$ gates corresponding to the two interaction vertices in the Feynman diagram in Fig.~\ref{fig:quarkselfenergy}.
We emphasise that $Q$ does not act on the $\tilde{q}$ register.
We see from eq.~\eqref{eq:Q_gate_behaviour} that after applying the $Q$ gate once, the state of the quantum computer becomes
\begin{equation}
\frac{1}{\sqrt{24}} \sum_{
\substack{
a \in \{1,\ldots,8\}\\
j,k \in \{1,2,3\}
}
} T^a_{jk} \ket{a}_g\ket{j}_q\ket{k}_{\tilde{q}}\ket{\Omega}_\mathcal{U} + \left(\textrm{terms orthogonal to } \ket{\Omega}_\mathcal{U} \right)
\end{equation}
and after applying the second $Q$ gate, the state becomes
\begin{equation}\label{eq:example_circuit_after_second_Q_gate}
\frac{1}{\sqrt{24}} \sum_{
\substack{
a \in \{1,\ldots,8\}\\
i,j,k \in \{1,2,3\}
}
} T^a_{ij}T^a_{jk} \ket{a}_g\ket{i}_q\ket{k}_{\tilde{q}}\ket{\Omega}_\mathcal{U} + \left(\textrm{terms orthogonal to } \ket{\Omega}_\mathcal{U} \right) .
\end{equation}
This looks somewhat like the desired colour factor, but it is not immediately accessible.
In particular, the state contains a sum over $a$ but each term $T^a_{ij}T^a_{jk}$ multiplies a distinct state $\ket{a}_g$ of the gluon register, which means that the desired summation $\sum_a T^a_{ij}T^a_{jk}$ has not yet been performed.

In order to perform the sum, we first observe by inverting eq.~(\ref{eq:Rg_effect}) that $R_g^{-1}$ acting on any state $\sum_{a=1}^8 c_a \ket{a}_g$ of the gluon register would produce the state
\begin{equation}
R_g^{-1} \sum_{a=1}^8 c_a \ket{a}_g = \left(\frac{1}{\sqrt{8}} \sum_{a=1}^8 c_a\right)\ket{\Omega}_g + \left(\textrm{terms orthogonal to } \ket{\Omega}_g \right) ,
\end{equation}
effectively averaging over the coefficients of the 8 colour states $\ket{a}_g$.
Similarly, it can be seen by inverting eq.~(\ref{eq:Rq_effect}) that $R_q^{-1}$ acting on any state $\sum_{i,k\in\{1,2,3\}} c_{ik} \ket{i}_q \ket{k}_{\tilde{q}}$ of the $q$ and $\tilde{q}$ registers would produce the state
\begin{equation}
R_q^{-1} \sum_{i,k\in\{1,2,3\}} c_{ik} \ket{i}_q \ket{k}_{\tilde{q}} = \left(\frac{1}{\sqrt{3}} \sum_{i=1}^3 c_{ii} \right)\ket{\Omega}_q \ket{\Omega}_{\tilde{q}} + \left(\textrm{terms orthogonal to } \ket{\Omega}_q\ket{\Omega}_{\tilde{q}} \right) ,
\end{equation}
effectively performing a trace over quark colours.
Note that tracing over external colours is not essential, but we have chosen to do so in order to allow each Feynman diagram to be validated by comparing a single number to the output of our quantum circuits.

Thus, after applying the $R_g^{-1}$ and $R_q^{-1}$ gates to the state produced in eq.~\eqref{eq:example_circuit_after_second_Q_gate}, we obtain the state
\begin{equation}\label{eq:example_trace_final_state}
\frac{1}{24}\left(
\sum_{
\substack{
a \in \{1, ..., 8\}\\
i,j \in \{1,2,3\}
}
} T^a_{ij} T^a_{ji}
\right)
\ket{\Omega}_g\ket{\Omega}_q\ket{\Omega}_{\tilde{q}}\ket{\Omega}_\mathcal{U} + \left(\textrm{terms orthogonal to } \ket{\Omega}_g\ket{\Omega}_q\ket{\Omega}_{\tilde{q}}\ket{\Omega}_\mathcal{U} \right).
\end{equation}
It can be observed in this state that the coefficient of the original reference state $\ket{\Omega}_g\ket{\Omega}_q\ket{\Omega}_{\tilde{q}}\ket{\Omega}_\mathcal{U}$ encodes the colour factor~\eqref{eq:self_energy_colour_factor} of the diagram.
This result can be generalised to arbitrarily more complicated diagrams by adding more qubits and more $Q$ and $G$ gates, as will be explained in the next section.
We note that the procedure described in this example can be applied at the level of either an unsquared diagram or a squared diagram, since for colour calculations the Feynman rules remain the same in both cases.

\section{Calculating the colour factor of arbitrary Feynman diagrams}\label{sec:generalisation}
It is straight-forward to generalise the illustrative example from sec.~\ref{sec:illustrative_example} to now calculate colour factors for Feynman diagrams with arbitrary numbers of quarks and gluons.
Given an arbitrary Feynman diagram with $N_q$ quark lines and $N_g$ gluons, the procedure is as follows:
\begin{enumerate}
\item Create a quantum circuit with a 3-qubit gluon register $g$ for each gluon, a pair of 2-qubit quark registers $q$, $\tilde{q}$ for each quark line, and a single unitarisation register $\mathcal{U}$.
\item Initialise each register $r$ to a reference state $\ket{\Omega}_r$ in which each qubit is in the state $\ket{0}$.
\item For each gluon, apply $R_g$ to the corresponding register $g$.
\item For each quark line, apply $R_q$ to the corresponding pair of registers $q,\tilde{q}$.
\item For each quark-gluon vertex, apply a $Q$ gate to the corresponding registers $g$ and $q$.
\item For each triple-gluon vertex, apply a $G$ gate to the 3 corresponding $g$ registers.
\item For each gluon, apply $R_g^{-1}$ to the corresponding gluon register.
\item For each quark, apply $R_q^{-1}$ to the corresponding pair of quark registers $q,\tilde{q}$.
\end{enumerate}
Just as with the illustrative example in sec.~\ref{sec:illustrative_example}, the colour factor $\mathcal{C}$ for the diagram is now found encoded in the final state of the quantum computer, which is
\begin{equation}\label{eq:final_state}
\frac{1}{\mathcal{N}} \mathcal{C} \ket{\Omega}_{all} + \left(\textrm{terms orthogonal to} \ket{\Omega}_{all}\right) ,
\end{equation}
where $\mathcal{N} = N_c^{n_q} \left(N_c^2-1\right)^{n_g}$ and
\begin{equation}
\ket{\Omega}_{all} = \left( \prod_{m=1}^{n_g} \ket{\Omega}_{g_m} \right) \left( \prod_{l=1}^{n_q} \ket{\Omega}_{q_l}\ket{\Omega}_{\tilde{q}_l} \right) \ket{\Omega}_{\mathcal{U}}.
\end{equation}

\section{Validation}\label{sec:validation}
To validate our methods, we implemented our circuits in \textsc{Python} using the \textsc{IBM Qiskit} framework.
We used this to run our circuits on a simulated noiseless quantum computer.
To verify that the state~\eqref{eq:final_state} is indeed being correctly produced, we ran each simulation $10^8$ times, measuring the final state each time, and inferred the colour factor $\mathcal{C}$ from the fraction of times that the output state was measured to be $\ket{\Omega}_{all}$.
While this is a simple and transparent way to verify the state~\eqref{eq:final_state}, it is not the most efficient way and we emphasise that more sophisticated measurement schemes are possible such as quantum amplitude estimation~\cite{Brassard:2000,Grinko:2019,Suzuki:2019,Nakaji:2020}, which offers a quadratic speed-up.
Nonetheless, it can be seen from the results in Table~\ref{tab:checks} that the measurements are fully consistent with the analytical expectation of the colour factors.

\begin{table}[p]
  \begin{center}
    \begin{tabular}{c|c|c}
    Diagram & Analytical & Numerical \\
    \midrule
    \thead{\includegraphics[width=0.3\textwidth]{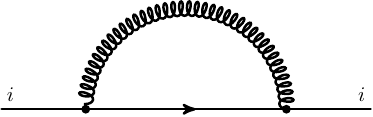}} & $C_F N = 4$ & $3.9988 \pm 0.0012$ %2776113/100000000 Events
    \\
    \midrule
    \thead{\includegraphics[width=0.3\textwidth]{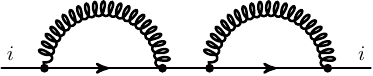}} & ${C_F}^2 N = \frac{16}{3}$ & $5.331 \pm 0.010$ %77082/100000000 Events
    \\
    \midrule
    \thead{\includegraphics[width=0.3\textwidth]{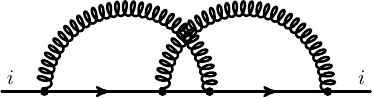}} & $\frac{C_F}{2} = \frac23$ & $0.673 \pm 0.010$ %1229/100000000 Events
    \\
    \midrule
    \thead{\includegraphics[width=0.3\textwidth]{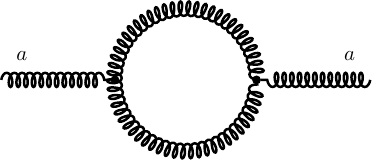}} & $N(N^2-1) = 24$ & $23.95 \pm 0.03$ %218753/100000000 Events
    \\
    \midrule
    \thead{\includegraphics[width=0.3\textwidth]{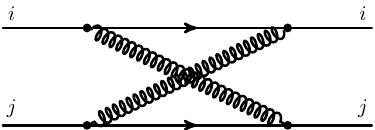}} & $\frac{(N^2-1)}{4} = 2$ & $2.00 \pm 0.03$ %1206/100000000 Events
    \\
    \midrule
    \thead{\includegraphics[width=0.3\textwidth]{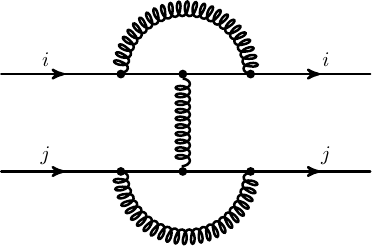}} & $0$ & $0.0^{+0.5}_{-0.0}$ %0/100000000 Events
    \\
    \midrule
    \thead{\includegraphics[width=0.3\textwidth]{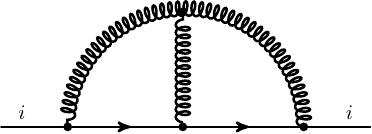}} & $\frac{C_F N^2}{2} = 6$ & $5.92 \pm 0.08$ %1486/100000000 Events
    \\
    \end{tabular}
  \end{center}
  \caption{\label{tab:checks}
    Colour factors for example Feynman diagrams.
    The first column depicts the Feynman diagrams, with indices on external legs indicating identical colours.
    The central column states the analytical result for the colour factor.
    The last column displays the numerical result for each colour factor obtained using quantum simulations in the manner explained in sec.~\ref{sec:validation} of the text.
    }
\end{table}

\section{Summary and Outlook}
In these proceedings, based on our article~\cite{Chawdhry:2023jks} in which full details can be found, we have designed quantum circuits to simulate the colours parts of perturbative QCD.
As an example application, we have shown how they can be used for calculating the colour factors of arbitrary Feynman diagrams.
This is a first step towards a full quantum simulation of perturbative QCD processes.

Our work opens up several natural avenues for further exploration.
Firstly, there is the interference of multiple Feynman diagrams. Quantum computers are naturally well-suited to this task, due to their ability to coherently manipulate quantum states, and we therefore believe that this extension should be straight-forward.
Secondly, one can try to implement the kinematic parts of Feynman diagrams.
This could re-use some of the ideas from this work, particularly the unitarisation register for implementing non-unitary operations.
However, one will also require methods to handle the much larger Hilbert space associated with kinematics.
Thirdly, one can eventually seek to combine these components into a quantum computer-based Monte Carlo simulation of cross-sections in order to obtain a quadratic speed-up over classical Monte Carlo simulations.

\section*{Acknowledgements}
The authors are grateful to Fabrizio Caola, Stefano Gogioso, Michele Grossi, and Joseph Tooby-Smith for helpful discussions.
The research of H.C.\ is supported by ERC Starting Grant 804394 \textsc{hip}QCD.
M.P.\ acknowledges support by the German Research Foundation (DFG) through the Research Training Group RTG2044.
H.C. is grateful to the Galileo Galilei Institute for hospitality and support during the
scientific program on ``Theory Challenges in the Precision Era of the Large Hadron
Collider,'' where part of these this proceedings paper was written.

\end{document}